\documentclass[%
 aip,
 amsmath,amssymb,
reprint
]{revtex4-1}

\usepackage{graphicx}
\usepackage{dcolumn}
\usepackage{xcolor}
\usepackage{bm}
\usepackage{xspace}
\newcommand{\NO}{nOb\xspace}
\newcommand{\NOs}{nObs\xspace}

\usepackage[utf8]{inputenc}
\usepackage[T1]{fontenc}
\usepackage{mathptmx}
\usepackage{etoolbox}

\usepackage[bookmarks,colorlinks,breaklinks]{hyperref}
\definecolor{dullmagenta}{rgb}{0.4,0,0.4}   
\definecolor{darkblue}{rgb}{0,0,0.4}
\definecolor{darkgreen}{rgb}{0,0.4,0}
\hypersetup{linkcolor=blue,citecolor=darkblue,filecolor=dullmagenta,urlcolor=darkgreen} 

\definecolor{niceblue}{rgb}{0.3,0.2,1.0}
\renewcommand{\revised}[1]{#1}

\makeatletter
\def\@email#1#2{%
 \endgroup
 \patchcmd{\titleblock@produce}
  {\frontmatter@RRAPformat}
  {\frontmatter@RRAPformat{\produce@RRAP{*#1\href{mailto:#2}{#2}}}\frontmatter@RRAPformat}
  {}{}
}%
\makeatother
\begin{document}
\sloppy

\title[\protect\sffamily Nano-object interferometry]{\protect\large\sffamily\bfseries
Internal decoherence in nano-object interferometry due to phonons\vspace*{1ex}}
\author{C. Henkel}%
\affiliation{
Universit\"at Potsdam, Institut f\"ur Physik und Astronomie, 14476 Potsdam, Germany%
}%
\email{henkel@uni-potsdam.de}

\author{R. Folman\vspace*{0.5ex}}
\affiliation{
Ben-Gurion University of the Negev, Department of Physics, Beer Sheva 84105, Israel%
}
\altaffiliation[Also at ]{Ilse Katz Institute for Nanoscale Science and Technology, Beer Sheva, Israel}

\date{27 April 2022}

\begin{abstract}
We discuss the coherent splitting and recombining of a nanoparticle in a mesoscopic ``closed-loop''  Stern-Gerlach interferometer in which the observable is the spin of a single impurity embedded in the particle. This spin, when interacting with a pulsed magnetic gradient, generates the force on the particle. We calculate the internal decoherence which arises as the displaced impurity excites internal degrees of freedom (phonons) that may provide \emph{Welcher Weg} information and preclude interference. We estimate the constraints this decoherence channel puts on future interference experiments with massive objects. {We find that for a wide range of masses, forces and temperatures, phonons do not inhibit Stern-Gerlach interferometry with micro-scale objects. However, phonons do constitute a fundamental limit on the splitting of larger macroscopic objects if the applied force induces phonons.}
\end{abstract}

\maketitle

\section{Introduction}
\label{sec:level1}

Quantum Mechanics (QM) and General Relativity (GR), the latter being the current theory of gravity, are the two pillars of modern physics. The quantum nature of gravity, or the unification of these two pillars, has been an open question of utmost importance for decades now. While theory has not been able to find a satisfactory solution to this question, it is of paramount importance for experiments to deliver hints. We discuss here a nano-object interferometer aimed at delivering such hints as they emerge from the interface between QM and GR. Such an interferometer may probe gravity-related ideas, from the mainstream quantum of gravitation\cite{Bronstein_1936}, named the graviton, to speculative ideas, such as those of Penrose concerning gravitationally induced collapse (see, e.g., Ref.\,\onlinecite{Marshall_2003})
or the short-range correction to gravity, the so-called fifth force. For an in-depth overview, see for example Refs.\,\onlinecite{Marshman_2020,Wan_2016,Marletto_2017,Bose_2017,Marshman_2020b,Marletto_2020,Carney_2021}.

Another specific motivation for such an interferometer relates to the foundations of QM. It will push the limits to which the accuracy of QM is tested, by several orders of magnitude: both in the amount of mass that is being put in a spatial superposition, and in the size of the spatial splitting. Another contribution to the foundations of QM would be the ability to test continuous spontaneous localization models. These are of crucial importance in the search for extensions to QM. For a more complete review of underlying concepts, see for example Refs.\,{\onlinecite{Romero-Isart_2011,Gasbarri_2021}}.

Finally, a third motivation is quantum technology. In addition to the quantum computer, there is metrology. A nano-object interference experiment will bring about cutting-edge metrological capabilities, e.g., in the measurement of gravity \cite{Rademacher_2020}, including geodesic studies and mineral searches, or in acceleration sensing.

Let us briefly note that the aims of such an endeavor are {difficult to be obtained} by laser-based matter interferometry. First, laser pulses can be absorbed and scattered by the nano-object (\NO) and, as already shown in experiments, they heat up the object and can lead to its {opaqueness} or destruction.\cite{Rahman_2016} Light scattering also increases the spatial decoherence rate, {if it can resolve the two paths}. Second, laser-based matter-wave interferometry requires an appropriate optical transition, and this severely restricts its applicability to massive solid objects. {Such optical transitions have been suggested \cite{Albrecht_2014}, but they suffer from two major drawbacks: (a) they typically require cryogenic temperatures and as noted the light impinging on the object heats it and cooling a free-space object is extremely hard; (b) they give very low recoil velocities and thousands of these transitions would be required for significant splitting in a short time. For example, for a $10^{-20}$\,kg \NO, the optical recoil velocity is on the order of $10^{-8}$\,m/s whereas a continued acceleration of say $100\,{\rm m/s}^2$ for a short period of say $100\,\mu$s gives a velocity six orders of magnitude higher.}
Here we consider a Stern-Gerlach interferometer (SGI) {which utilizes magnetic gradients rather than} light pulses to generate the spatial superposition.

The state of the art for \NO interferometry utilises a beam of heavy molecules impinging on a grating. The heaviest molecules put in a superposition to date consist of up to 2000 atoms.\cite{Fein_2019,Shayeghi_2020} The SGI may go far beyond this limit.
In the spirit of molecular-beam experiments, several proposals exist for slit-type experiments with higher masses, specifically including solid \NOs.\cite{Romero-Isart_2011,Pino_2018,Romero-Isart_2017} In these proposed experiments, none of which has been realized yet, the signal is a spatial interference pattern, reminiscent of the double-slit interference pattern. These proposals face several challenges. For example, for spatial interference patterns to form, a long time-of-flight (TOF) is required, and as the decoherence rate of delocalized massive-object states is expected to be high, a prolonged TOF seems impractical.
Furthermore, for many of these configurations, the periodicity of the interference pattern is expected to be extremely small, so that high spatial resolution is required for detection, a resolution which may be beyond available technology, especially for high-efficiency detection.

\begin{figure}[tbh]
\centerline{%
\includegraphics*[width=0.42\textwidth]{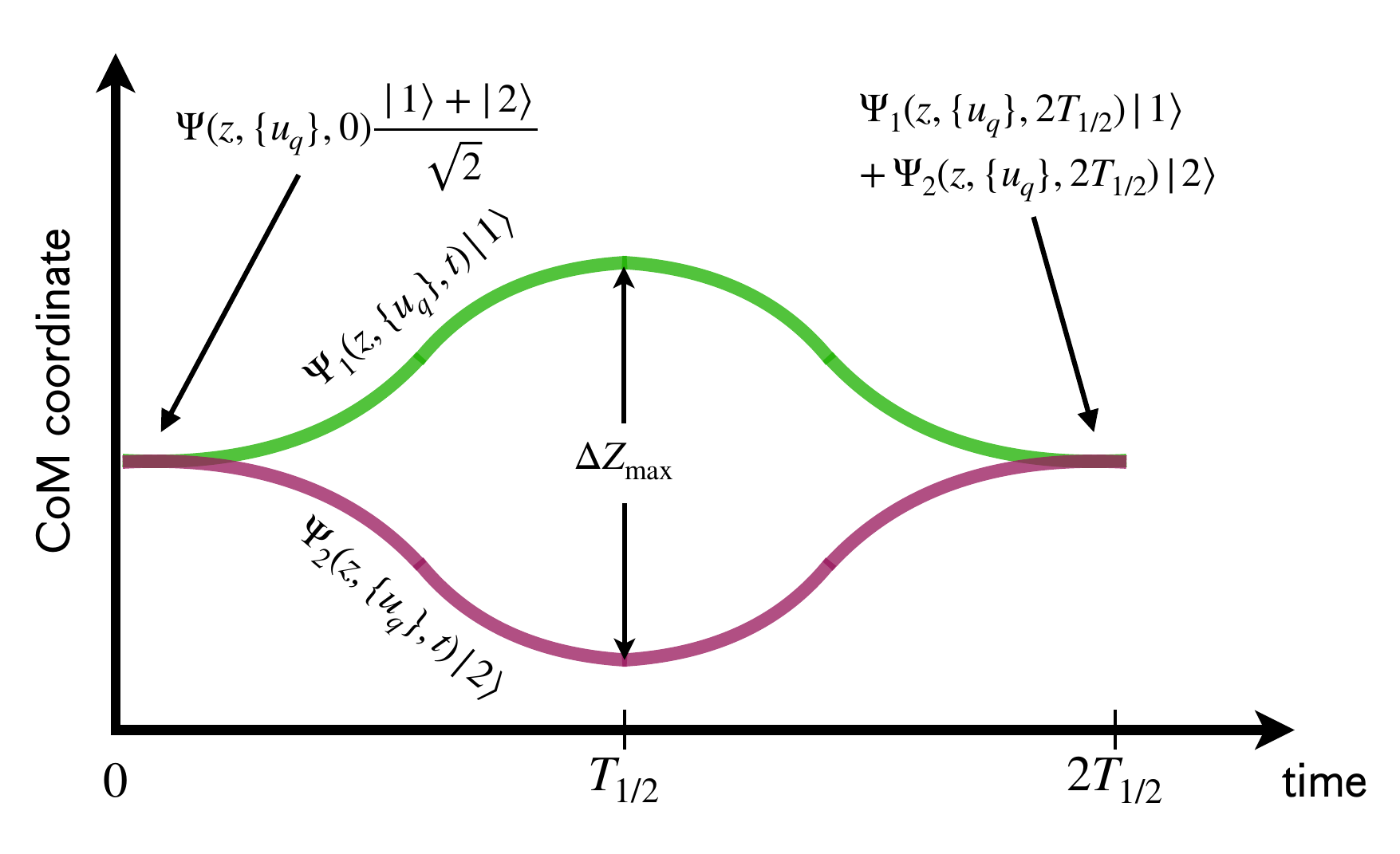}
}
\vspace*{-3ex}
\caption[]{%
Sketch of the Stern-Gerlach interferometer (SGI), adapted from Fig.\,1 in Ref.\,\onlinecite{Margalit_2021}.
Along the two paths, the wavepackets behave distinctly different with respect to the center-of-mass coordinate (CoM) $z$ and the amplitudes $\{u_q\}$ of the internal phonon modes.
The maximum splitting $\Delta Z_{\rm max}$ is reached {at $T_{1/2}$ which is} half the time required to close the loop.
\\
\revised{Adapted with permission from Y. Margalit, O. Dobkowski, Z. Zhou, O. Amit, Y. Japha, S. Moukouri, D. Rohrlich, A. Mazumdar, S. Bose, C. Henkel, and R. Folman, ``Realization of a complete Stern-Gerlach interferometer: Towards a test of quantum gravity,'' {\it Science Adv.} {\bf 7}, eabg2879 (2021). Copyright (2021) The Authors under licence CC BY 4.0, DOI:10.1126/sciadv.abg2879.}
}
\label{fig:sketch-ifm}
\end{figure}

The SGI considered here is a completely different route for realizing such an interferometer, as shown in Fig.\,\ref{fig:sketch-ifm}. A single spin embedded in the nano-object is first put in a superposition of opposite spins. When moving through a magnetic-field gradient, the two spin orientations experience opposite forces, and this splits the wavefunction of the entire \NO into two wave-packets (WPs), effectively generating spin-momentum entanglement.
Reversing the opposite forces is then used to bring the two WPs back together again. Such an interferometer has three crucial advantages: (a) the splitting is active, namely, a real force is utilized, and it does not depend on expanding the WP; consequently, large splitting distances may be achieved in a short time; (b) the signal forms independently of any TOF and, as the recombination is also active, it may be done in a short time;
(c) the observable is spin population, as in modern atom interferometry; in contrast to spatial fringes, its detection does not require high spatial resolution.

We recently demonstrated the coherence of a Stern-Gerlach ``closed-loop'' interferometer with ultracold atoms and discussed the possibility of realizing such an interferometer with a nano-particle.\cite{Margalit_2021}
{Decoherence, or lack of coherence, comes from two very different physical processes, the first originating in the coupling of the \NO to the environment, and the second originating in the fact that the loop is not completely closed, coined the ``Humpty-Dumpty'' effect (see Ref.\,\onlinecite{Margalit_2021} and references therein). The latter type of decoherence, independent of the environment, depends on the final overlap of several wave functions, that of the external degrees of freedom (position, momentum and rotation, for the latter, see for example Ref.\,\onlinecite{Japha_2022} and references therein), as well as internal degrees of freedom, including spin and phonons, the latter being the topic of this work. If the wave functions of the WPs along the two paths become orthogonal, no interference is possible. Orthogonality may arise from the non-identical phonon excitations due to the applied Stern-Gerlach force along the two paths the WPs take. As we show in the following, this is dependent on the \NO size and temperature and on the applied force (i.e., the magnetic gradient).

As we would like to focus solely on decoherence due to phonons, in the following we assume that all other types of decoherence are negligible. However, let us briefly note the different types of decoherence which should be considered, and the works which have already addressed them. To begin with, spin coherence is crucial if spin is the observable of the interferometer. The $T_1$ and $T_2$ times of the spin must be longer than the interferometer time. For the nitrogen-vacancy center embedded in a nano-diamond as discussed in this work, a room-temperature coherence time of a few hundred $\mu$s was observed.\cite{Trusheim_2013} Blackbody radiation (BBR) is a source of spatial decoherence.\cite{Romero-Isart_2011,Bateman_2014,Chang_2010,Kamp_2020,Schut_2022} Collisions with background gas are also an important source of decoherence.\cite{Albrecht_2014,Kamp_2020} Any charging of the \NO would again cause decoherence through the electrostatic coupling to the environment.\cite{Sonnentag_2007} See also Ref.\,\onlinecite{Margalit_2021} and references therein.}

Many works have already considered {the importance of internal temperatures for lower decoherence in such \NO interferometers.} Specifically, concerning BBR, it has been suggested that internal state cooling of the neutral test masses in addition to external cooling in opto-mechanical cavities would prove to be greatly beneficial.\cite{Xuereb_2013} Furthermore, quite a few works dealt with the quantum dynamics of \NO phonons.\cite{Huemmer_2020, GonzalezBallestero_2020b,GonzalezBallestero_2020a}
However, as far as we know, this is the first treatment of phonons as a source of decoherence.

\section{Model}
\label{s:model}

\subsection{Hamiltonian}
\label{sec:level2}

We envision a nano-object embedded with a single spin, such as a nano-diamond with a single nitrogen-vacancy center.
The model for the nano-object is based on microscopic models of
magnetic materials, see, e.g., Ref.\,\onlinecite{Strungaru_2021}.
The atoms of
the object are characterized by their positions ${\bf r}_i$ and momenta
${\bf p}_i$. For our purposes, only a single atom, {say at site number $s$,} carries
a spin ${\bf S}$. For definiteness we assume $S=\frac{1}{2}$. The Hamiltonian is a sum of mechanical and
magnetic terms $H = H_1 + H_2(t)$ with
\begin{align}
H_1 &= \sum_{i} \frac{ {\bf p}_i^2 }{ 2 m } + \sum_{\langle i,j \rangle}
V( {\bf r}_i - {\bf r}_j )
\label{eq:energy-of-chain}
\\
H_2(t) &= - \mu {\bf S} \cdot {\bf B}({\bf r}_s, t)
\end{align}
where $m$ is the single-atom mass (assumed identical for simplicity) and
$\mu$ the magnetic moment (the spin ${\bf S}$ is taken
dimensionless). {We do not take into account collective magnetic interactions, such as the diamagnetic interaction \cite{Pedernales_2020, Marshman_2022}, as they are not expected to excite phonons.}

The second sum in Eq.\,(\ref{eq:energy-of-chain}) can be restricted to nearest neighbor sites
$i,j$ and involves the bond potential
\begin{equation}
V({\bf r}_i - {\bf r}_j) = \frac{ K }{ 2 } ({\bf r}_i - {\bf r}_j)^2 + \text{anharmonic terms}
\label{eq:spring-potential}
\end{equation}
with a common spring constant $K$.
From this model, we get for the center-of-mass coordinate
${\bf R} = (1/N) \sum_i {\bf r}_i$ and its corresponding momentum ${\bf P} = \sum_i {\bf p}_i$
the equation of motion
\begin{equation}
\dot {\bf P} = - \sum_i \frac{ \partial H }{ \partial {\bf r}_i }
= \mu \sum_\alpha {S}_\alpha \frac{ \partial B_\alpha }{ \partial {\bf r}_s }
\label{eq:center-of-mass}
\end{equation}
where the interaction potential Eq.\,(\ref{eq:spring-potential}) drops out by Newton's \emph{actio} = \emph{reactio}. Typical accelerations are given in Table \ref{tab:parameters}.

\begin{table}
\caption[]{Typical orders of magnitude for nano-particles made from diamond.}
\label{tab:parameters}
\begin{ruledtabular}
\begin{tabular}{lll}
magnetic gradient\footnote{%
Achievable with atom chips $1\,\mu$m away from a $1\times 1 \mu$m$^2$ wire
with $10^9\,$A/cm$^2$ current density.}
& $10^{6}\,{\rm T/m}$
\\
particle mass
& $10^{6}-10^{10}\,{\rm amu}$
\\
size (diameter)%
\footnote{%
Unit cell with $a = 3.6\,$\AA{} and $8$ C atoms.}
&
$L = 10-200\,{\rm nm}$
\\
magnetic moment $\mu$ & $1\,\mu_{\rm B} = h \times 14\,{\rm GHz}/{\rm T}$
\\
acceleration & $6000-0.6\,{\rm m/s}^2$
\\
lowest phonon mode\footnote{%
Speed of sound $c = 17.5\,{\rm km/s}$.
}
\\
$\omega_1/2\pi = c/2L$
& $900-45\,{\rm GHz}$
\\
\kern-0.5ex
\begin{tabular}{l}
de Broglie wavelength\footnote{%
Estimated as $\lambda_{\rm cm} = \hbar (M k_{\rm B} T_{\rm cm})^{-1/2}$ with the
center-of-mass temperature $T_{\rm cm}$.
}
$\lambda_{\rm cm}$
\\
phonon coherence length\footnote{%
Estimated as $\lambda_{\rm ph} = \hbar (M k_{\rm B} T_{\rm ph})^{-1/2}$ with the
internal (phonon) temperature $T_{\rm ph}$.
}
$\lambda_{\rm ph}$
\end{tabular}
&
\kern-0.5ex
\begin{tabular}{ll}
$T_{\rm cm} = 293\,{\rm K}:$ & $10^{-14}-10^{-16}\,{\rm m}$
\\
$T_{\rm ph} = 4\,{\rm K}:$ & $10^{-13}-10^{-15}\,{\rm m}$
\end{tabular}
\end{tabular}
\end{ruledtabular}
\end{table}

\begin{figure}[htbp]
   \includegraphics[width=0.3\textwidth]{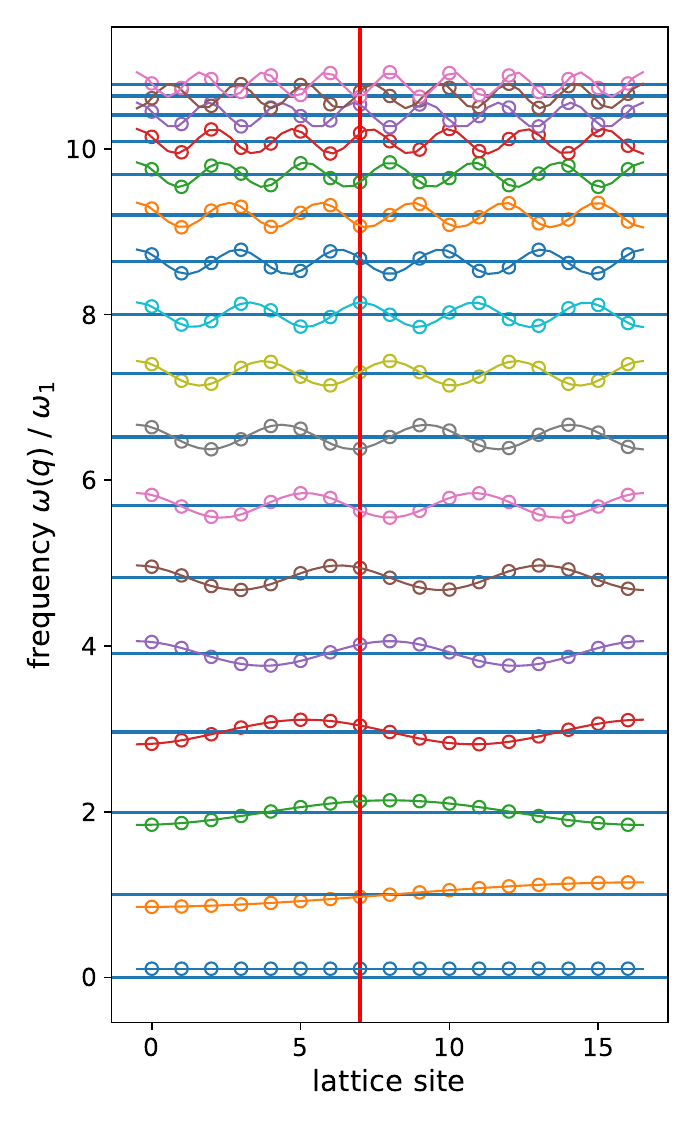}
\vspace*{-3ex}
   \caption[]{Sketch of phonon modes (standing waves) for a linear chain.
   The total length
   is covered by an integer multiple of half the wavelength.
   The vertical red line marks
   the position of the impurity spin. Frequencies scaled to the fundamental
   tone $\omega_1 = \pi c / L$ with the speed of sound $c$ and length
   $L = 17$~unit cells. The phonon amplitude is represented by the
   displacements of the circles, but note that in this linear model, this
   is actually a longitudinal phonon. The zero-frequency mode corresponds
   to a displacement of the chain as a whole (center of mass mode).}
   \label{fig:sketch-phonons}
\end{figure}

We start for the phonons with a simple one-dimensional model with $N$
atoms in a linear chain having a total mass of $m N = M$.
If we interpret the coordinates $z_i$ as the deviation from equilibrium
positions spaced by the equilibrium bond length (lattice constant) $a$,
we get a chain model whose phonon spectrum in the harmonic approximation
is given by (force-free or Neumann boundary conditions)
\begin{equation}
\omega(q) = \sqrt{ \frac{ 4 K }{ m } } \sin \frac{q a}{2}
\,,\qquad
q = \frac{ \pi }{ L }\left\{ 0, 1, 2, \ldots, N_x - 1\right\}
\label{eq:phonon-dispersion}
\end{equation}
where $L = a N$ is the total length of the chain.
The mode amplitudes $u_q$ (illustrated in Fig.\,\ref{fig:sketch-phonons})
can be computed from the projection
\begin{equation}
u_q = \frac{ 2 }{ N } \sum_n z_n \cos[(n+\tfrac12) qa]
\,, \quad
q \ne 0
\label{eq:mode-uq}
\end{equation}
and for them, we get the equations of motion
\begin{equation}
\frac{ {\rm d}^2 u_q }{ {\rm d}t^2 } =
- \omega^2(q) u_q
+ \frac{ 2 \mu }{ M }
\cos[(s+\tfrac12)qa]
\sum_\alpha S_\alpha
\frac{ \partial B_\alpha }{ \partial z_s }
\label{eq:phonon-mode-q}
\end{equation}
where the integer $s$ labels the equilibrium position of the impurity spin (the left end of the chain is at $s = 0$).%
\footnote{We checked that in the limit $q \to 0$, one recovers the center-of-mass
dynamics of Eq.\,(\ref{eq:center-of-mass}).}
Finally, the spin itself precesses according to the Larmor equation
\begin{equation}
\frac{ {\rm d}{\bf S} }{ {\rm d}t } = \frac{\mu}{\hbar} {\bf S} \times
{\bf B}( {\bf r}_s, t )
\label{eq:spin-precession}
\end{equation}

We consider here the simplest setting where the magnetic field is aligned to the $z$-axis and is a linear function 
$B_z(z, t) = B_0 + b(t) z$.
\revised{Indeed, this is the experimental configuration used by the atomic SGIs \cite{Margalit_2021}. The linear field significantly simplifies the dynamics, as there is no coupling of the CoM, spin and phonon degrees of freedom. This is valid provided the ratio $|b(t) {\bf R} / B_0|$ remains small enough to neglect the contributions from transverse components of the magnetic field (required by $\mathop{\rm div} {\bf B} = 0$).}

In the ``spin up'' or ``down'' configurations, the spin vector ${\bf S}(t) =
\pm \tfrac12 {\bf e}_z$ will thus be stationary (no precession, no spin flips).
The Stern-Gerlach force becomes independent of the nano-object position,
and we
may solve Eqs.\,(\ref{eq:center-of-mass}, \ref{eq:phonon-mode-q}) easily:
\begin{eqnarray}
P(t) &=& P(0) \pm \frac{ \mu }{ 2 } \int_0^t\!{\rm d}t'\, b(t')
\label{eq:cm-displacements}
\\
Z(t) &=& Z(0) + \frac{1}{M} \int_0^t\!{\rm d}t' \, P(t')
\\
u_q(t) &=& u_q(0) \cos \omega_q t
+ \frac{ \dot u_q(0) }{ \omega_q } \sin \omega_q t
\label{eq:solutions-P-and-uq}
\\
&& {} \pm \frac{ \mu }{ M \omega_q }
\cos[(s+\tfrac12)qa]
\int_0^{t}\!{\rm d}t'\,
b(t') \sin[\omega_q(t-t')]
\nonumber
\end{eqnarray}
The $\pm$ sign of the spin projection thus determines the direction of the momentum,
as well as the sign of the phonon amplitude.
The equations of motion being linear, these
expressions are {equally valid, whether the variables are treated classically or as quantum operators (Heisenberg picture).}

\section{Overlap and Contrast}

The solution presented in Eqs.\,(\ref{eq:cm-displacements}--\ref{eq:solutions-P-and-uq}) determines
the contrast of the spin interference signal in the following way.
The key element is the overlap
between the wavepackets for the center-of-mass degree of freedom (DoF)
and the phonon DoFs that have evolved with either
sign $\pm$ of the spin projection.

\subsection{Center of mass}
\label{s:CoM-overlap}

To illustrate our method of calculation, we first calculate the overlap for the
center of mass (CoM). As noted in the introduction, we will eventually assume that it is perfect, so that loss of contrast is only due to orthogonality between phonon
excitations.

Consider for the CoM state an initial pure state
$|\psi\rangle$, and re-write the solution for momentum
$P(t)$ and position $Z(t)$ as the action of some displacement operator
$D_\pm$ where the spin sign $\pm$ corresponds to opposite displacement
directions:
\begin{align}
&\mbox{spin up:}\quad & P(t) &= D_+^\dag P(0) D_+
\nonumber\\
&& Z(t) &= D_+^\dag [ Z(0) + P(0)t / M ] D_+
\nonumber
\\
&\mbox{spin down:}\quad & P(t) &= D_-^\dag P(0) D_-
\nonumber\\
&& Z(t) &= D_-^\dag [ Z(0) + P(0)t / M ] D_-
\label{eq:}
\end{align}
The overlap in question can then be written as
\begin{equation}
\mathop{\rm Tr} \big(
D_+ |\psi\rangle \langle \psi| D_-^\dagger \big)
=
\langle \psi | D_-^\dagger
D_+ |\psi\rangle
\label{eq:overlap-cm}
\end{equation}
This goes down to zero contrast in a Gaussian fashion when the splitting exceeds the
width of the initial state $|\psi\rangle$, as expected for orthogonal states.

Since the inverse displacement $D_-^\dagger$ is the same as $D_+$,
the displacement operators
can be combined into one operator $D$
that
depends on the relative splittings $\Delta Z(t)$, $\Delta P(t)$
between the spin-down and spin-up trajectories.
From this viewpoint, the overlap may be understood as the amplitude to stay
in the initial state $|\psi\rangle$ after applying the operator $D$. Such
an amplitude is known in scattering as the Debye-Waller factor
(see, e.g., Ref.\,\onlinecite{Lipkin_1961,Levi_1979}),
and may also be related to the zero-phonon line in molecular spectroscopy
\cite{Sild_1988,Hsu_1984}.

It is now a well-known identity (sometimes called
the Bloch formula) that the expectation value of $D$ is related to the
Wigner representation $W(z, p)$
of the wave function $|\psi\rangle$, notably its double Fourier transform \cite{Schleich_book, VogelWelsch_Book, Milonni_book, Friesch_2000}
\begin{equation}
\langle \psi | D | \psi \rangle =
\chi(k, s)
= \int\!{\rm d}x \, {\rm d}p\,
W(z, p) \, {\rm e}^{ {\rm i} (k z - s p) }
.
\label{eq:Wigner-transform}
\end{equation}
The latter is also known as characteristic function for symmetrically
ordered products. Its arguments correspond to the displacements in phase
space: $k = \Delta P(t)/\hbar$, $s = \Delta Z(t)/\hbar$ that can be read
off from Eq.\,(\ref{eq:cm-displacements}).

The value $\chi(0, 0) = 1$ corresponds to full contrast
when the displacements in position and momentum are exactly zero. This defines
the target conditions for
a ``closed loop'' in phase space. Any deviations from them therefore
characterize the accuracy that is needed to close the interferometer.
In interferometry, it is well known that a non-closed loop generates a
so-called separation phase proportional to $\Delta P(t)$ and $\Delta Z(t)$.%
\cite{Bongs_2006, AmitPhD} The expression~(\ref{eq:Wigner-transform})
corresponds to the average of the corresponding phase factor over the
initial position and momentum distribution.

The initial distribution is indeed likely to be nonpure, and the advantage of using the
Wigner characteristic function is that it can be carried
simply through, if in Eqs.\,(\ref{eq:overlap-cm}, \ref{eq:Wigner-transform})
we are dealing with a density operator $\rho$
rather than the pure state $|\psi\rangle$. If we assume for simplicity that
the initial Wigner function is a Gaussian with variances $\sigma_p^2 = M k_{\rm B} T_{\rm cm}$
(kinetic temperature $T_{\rm cm}$ for the center-of-mass DoF)
and $\sigma_z^2$, we get the contrast reduction factor
\begin{align}
C_{\rm cm} & =
\left| \mathop{\rm tr}( D \rho ) \right|
=
\left| \chi(\Delta P(t)/\hbar, \Delta Z(t)/\hbar) \right|
\nonumber\\
& =
\exp\left[ - \tfrac{1}{2} \Delta P(t)^2 \sigma_z^2 / \hbar^2
- \tfrac{1}{2} \Delta Z(t)^2 \sigma_p^2 / \hbar^2
 \right]
\label{eq:cm-Gaussian}
\end{align}
The characteristic (rms) width of this Gaussian in $\Delta Z(t)$ is given by the
spatial coherence length $\hbar / \sigma_p = \lambda_{\rm cm}$ that coincides
simply with the thermal de Broglie wavelength $\lambda_{\rm cm}$ of the \NO
(see Table~\ref{tab:parameters}).
This suggests formidable precision requirements for closing the phase-space loop. It turns out, however, that one may confine the \NO in some potential and cool its CoM motion down to the ground state, thereby increasing its coherence length. \cite{Deli_2020,Tebbenjohanns_2020,Tebbenjohanns_2021,Magrini_2021}

\subsection{Internal degrees of freedom (phonons)}

The decoherence due to phonons can be estimated in a similar way. Consider first a fixed phonon mode with frequency $\omega_q \ne 0$.
Along the two interferometer arms, the amplitude $u_q$ of this mode suffers a
differential displacement in phase space given by,
from Eq.\,(\ref{eq:solutions-P-and-uq}):
\begin{align}
\Delta u_q(t) &= \frac{ 2 \mu }{ M \omega_q }
\cos[(s+\tfrac12)qa]
\int_0^{t}\!{\rm d}t'\,
b(t') \sin[\omega_q(t-t')]
\nonumber
\\
\Delta \dot u_q(t) &= \frac{ 2 \mu }{ M }
\cos[(s+\tfrac12)qa]
\int_0^{t}\!{\rm d}t'\,
b(t') \cos[\omega_q(t-t')]
\label{eq:def-Delta-uq}
\end{align}
Now consider this mode to be initially in thermal equilibrium at the (internal) temperature $T_{\rm ph}$.
Its contribution to the
energy in Eq.\,(\ref{eq:energy-of-chain})
is $E_q = \tfrac{1}{4} M (\dot u_q^2 + \omega_q^2 u_q^2)$, the magnetic
interaction energy being an irrelevant constant in a slowly varying field.
The initial phonon mode amplitude
thus has a Gaussian
Wigner function with variances $\sigma_{u,q}^2 = 2 k_{\rm B} T_{\rm ph} / (M \omega_q^2)$ in
displacement
and $\sigma_{\dot u,q}^2 = 2k_{\rm B} T / M$ in the corresponding velocity.
These variances actually provide the classical limit only:
at low temperatures, the replacement
\begin{equation}
k_{\rm B} T_{\rm ph} \mapsto \tfrac12 \hbar \omega_q \coth\frac{ \beta \omega_q }{ 2 }
\,, \qquad
\beta = \frac{ \hbar }{ k_{\rm B} T_{\rm ph} }
\label{eq:quantum-limit-Wigner}
\end{equation}
applies to get the correct Wigner function.\cite{Wigner_1932}
The width in amplitude $\sigma_{u,q}$
at zero temperature for the lowest phonon mode is,
at realistic temperatures, comparable
to the thermal phonon coherence length $\lambda_{\rm ph}$
given in Table~\ref{tab:parameters}.

\begin{figure}[tbhp]
   \centering
   \includegraphics[width=0.35\textwidth]{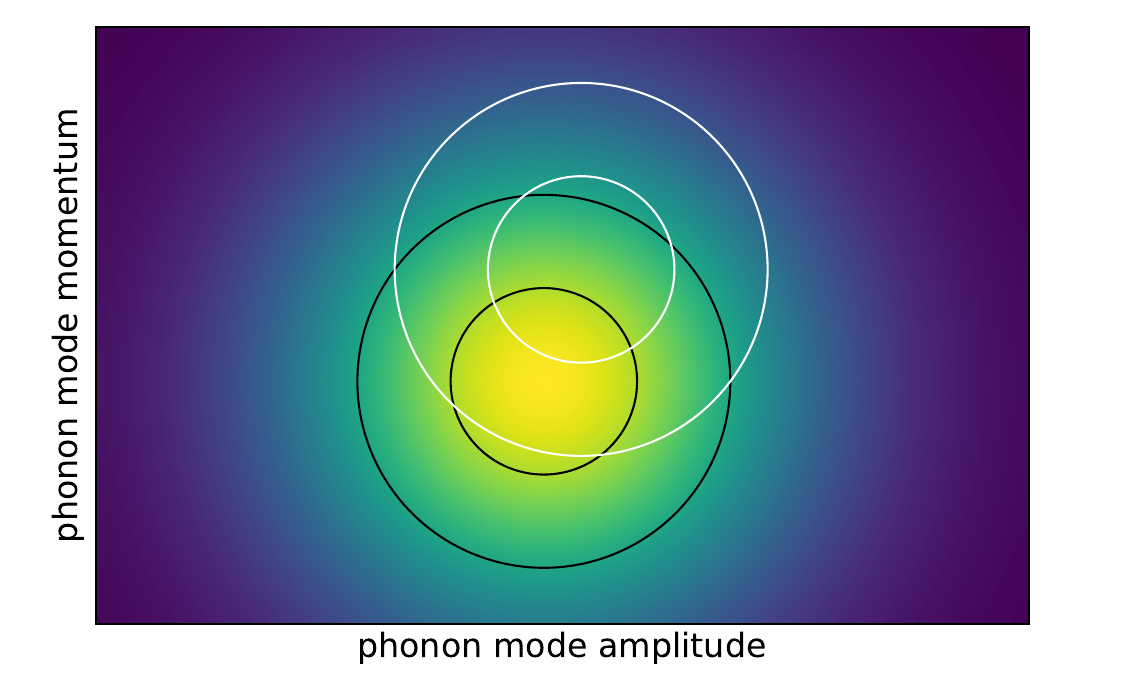}
\vspace*{-3ex}
   \caption[]{Splitting in {the phase space of a fixed phonon mode}. The contrast in the interferometer is
   determined by the overlap between the distributions illustrated by
   black contours (with color shading) and white contours.
   }
   \label{fig:phase-space-shift}
\end{figure}

This Wigner function is invariant under the rotation in phase space (first
line of Eq.\,(\ref{eq:solutions-P-and-uq})), this is why we may focus on
the displacements~(\ref{eq:def-Delta-uq}).
For the contrast reduction
due to phonon mode $q$, we need the variables $k_q, s_q$ in the displacement
operator $D_q = \exp[ {\rm i} (k_q u_q - s_q \dot u_q)]$. The
amplitude operators $u_q$ and $\dot u_q$ satisfy the commutation relations
\begin{equation}
[u_q, \dot u_{q'}] = \frac{ 2 {\rm i} \hbar }{ M } \delta_{qq'}
\label{eq:}
\end{equation}
that follow from Eq.\,(\ref{eq:mode-uq}), and we find
$k_q = M \Delta \dot u_q(t) / 2 \hbar$,
$s_q = M \Delta u_q(t) / 2 \hbar$.
The overlap for the mode $q$ thus generates a contrast
\begin{align}
C_q &= \exp\Big[ -
\tfrac12 k_q^2 \sigma^2_{u,q} + \tfrac12 s_q^2 \sigma^2_{\dot u,q}
\Big]
\nonumber\\
&= \exp \bigg[-
\frac{ M \omega_q }{ 8 \hbar } \left( \Delta u_q(t)^2 + \frac{ \Delta \dot u_q(t)^2 }{
\omega_q^2 } \right)
\coth\frac{ \beta \omega_q }{ 2 }
\bigg]
\label{eq:contrast-per-phonon-q}
\end{align}
We can here read off the characteristic phonon coherence ``length''
relative to which the splitting $\Delta u_q(t)$ of the phonon amplitude
between the two interferometer arms must be nullified.
For low-frequency modes
(i.e., $\hbar\omega_q \ll k_{\rm B} T_{\rm ph}$), this scale
is simply given by the thermal de Broglie wavelength with
$\lambda_{\rm ph} = \hbar / (M k_{\rm B} T_{\rm ph})^{1/2}$
(in the ${\rm pm}$ range or below, see Table~\ref{tab:parameters}).
The splitting in the phase space of phonon amplitude and
momentum can be visualized as in Fig.\,\ref{fig:phase-space-shift}. \revised{(A similar picture would represent the results obtained in Sec.\,\ref{s:CoM-overlap} for the CoM mode.)}
We also recall the alternative interpretation in terms of a Debye-Waller
factor: the contrast depends on the probability that the excitation of the mode
$u_q$ in one path of the WP relative to the other does \emph{not} differ by a single phonon quantum.

The two terms in the parenthesis of Eq.\,(\ref{eq:contrast-per-phonon-q})
can be combined into a complex, finite-window
Fourier transform. Introducing the time-dependent acceleration
$a(t) = \frac{ \mu }{ M } b(t)$ and the integral (with dimension velocity)
\begin{equation}
a(\omega_q, t) = \int_0^{t}\!{\rm d}t'\,
a(t') \,{\rm e}^{ {\rm i} \omega_q(t'-t) }
\label{eq:def-acceleration}
\end{equation}
we can write
\begin{equation}
\Delta u_q(t)^2 + \frac{ \Delta \dot u_q(t)^2 }{
\omega_q^2 }
= \frac{4}{\omega_q^2} \cos^2[(s+\tfrac12)qa] |a(\omega_q, t)|^2
\label{eq:overlap-per-mode}
\end{equation}

The extension to the full phonon spectrum is immediate if we assume that the
variables $\{u_q, \dot u_q \}$ for $q$ in the Brillouin zone
describe the normal modes of the chain of atoms.
In the harmonic approximation, this is obviously true, and the initial state
thus factorizes into a product of thermal states per normal mode. The overlaps per
mode multiply, and we get the phonon-based contrast reduction in the form
\begin{equation}
C_{\rm ph} = \exp\Big[ - \sum_q
\frac{ M }{ 2 \hbar \omega_q }
\coth\big(\tfrac12 \beta \omega_q\big)
\cos^2[(s+\tfrac12)qa]
|a(\omega_q, t)|^2
\Big]
\label{eq:result-phonon-contrast}
\end{equation}
A similar technique has been used for the dephasing of a qubit whose energy splitting couples to a phonon bath.\cite{VanKampen_1995,Palma_1996}

\begin{figure}[htb]
\centerline{%
\includegraphics*[width=0.47\textwidth]{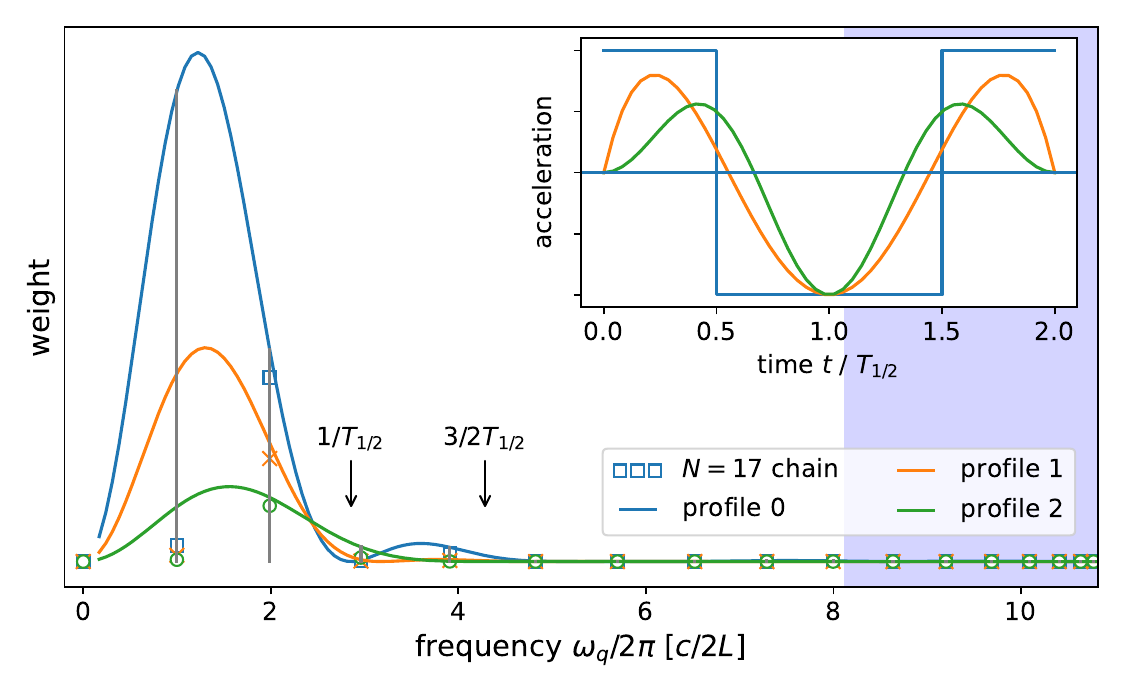}
}
\vspace*{-3ex}
\caption[]{%
Sketch of the contribution of individual phonon modes
to the interference contrast $C_{\rm ph}$. We plot
the terms under the sum in Eq.\,(\ref{eq:result-phonon-contrast})
as a function of phonon frequency, scaled to the fundamental tone
$\omega_1/2\pi = c / (2L)$.
The solid lines give an upper limit, discrete phonons are marked
by vertical lines, the symbols below the upper limit
({\footnotesize$\square$}, $\times$, $\circ$)
include the squared amplitude
of the phonon standing wave at the spin site $s = 7 \simeq N/2$.
In the indigo shading, $\hbar \omega_q \ge k_B T_{\rm ph}$
with the phonon temperature $T_{\rm ph}$.
Profiles 0, 1, 2: acceleration protocols
$a_0(t)$, $a_1(t)$, $a_2(t)$
of Eqs.\,(\ref{eq:simple-a0-model}--\ref{eq:simple-a2-model}),
the last one (green) having the smoothest switching-on 
and being the most adiabatic, see inset.
The width of the spectrum is inversely proportional to
the duration $T_{1/2}$ of the closed loop (arrows).
Parameters: chain with $N = 17$ atoms,
duration of applied forces compared to sound roundtrip $2T_{1/2} = 1.4\,L/c$, (internal) temperature $k_B T_{\rm ph} = 8.1\,\hbar\omega_1$.
}
\label{fig:sketch_summand}
\end{figure}

In the following section, we discuss the sum over all modes in
Eq.\,(\ref{eq:result-phonon-contrast}). To get a qualitative picture,
consider first the behaviour of the terms in the sum as a function of phonon
frequency $\omega_q$. A sketch is provided in Fig.\,\ref{fig:sketch_summand}.
The main feature is the Fourier spectrum
$a(\omega_q, t = 2T_{1/2})$
of the acceleration that
is nonzero in a range of frequencies $0 < \omega \lesssim 2\pi/T_{1/2}$.
Here,
the half duration $T_{1/2}$ of the interferometer loop corresponds to the
moment of maximum wavepacket splitting. The vertical lines with symbols
illustrate the positions
of the phonon eigenfrequencies, they are approximately harmonics of the
`fundamental  tone' $\omega_1/2\pi = c / 2 L$ where $c$ is the speed of sound.
The curves provide
an upper limit to the summands in Eq.\,(\ref{eq:result-phonon-contrast}),
the symbols give smaller values because they take into account the phonon mode amplitude
$\cos[(s+\tfrac12)qa]$. Impurity spins located near the \NO center present
two advantages: their overlap with the fundamental mode is small there (see
Fig.\,\ref{fig:sketch-phonons}), and their spin coherence time is maximal
because they avoid enhanced surface noise.\cite{Romach_2015}
Spins of nitrogen-vacancy (NV) centers in nano-diamonds have already exhibited a room-temperature coherence time of $200\,\mu{\rm s}$.\cite{Trusheim_2013} While it may be assumed that significant material engineering will improve these numbers \cite{Herbschleb_2019} (the state of the art for room-temperature bulk is $3\,{\rm ms}$ \cite{BarGill_2013}), even $200\,\mu{\rm s}$ is enough for a chip-based Stern-Gerlach interferometer to achieve significant splitting. Utilising well-known NV techniques (e.g., recent work of the Ben-Gurion University of the Negev group~\cite{Waxman_2014,Schlussel_2018,Rosenzweig_2018}), we do not see any fundamental spin-related obstacles.

\subsection{Discrete phonon spectrum (small objects)}

The object is small if its fundamental tone is much higher than the inverse
duration of the splitting pulse, i.e., $\omega_1 T_{1/2} \gg 2\pi$.
With $T_{1/2} = 30\,\mu{\rm s}$
and $c \approx 20\,{\rm km/s}$ (diamond), this applies for objects with
$L < 30\,{\rm cm}$, i.e. for any realistic small particle.
The amount of
orthogonality `hidden' in the phonon amplitudes along the spin up and down
paths of the wavepacket is then determined by the tails of the Fourier
spectrum $a(\omega, 2T_{1/2})$ of the \NO acceleration
(see Fig.\,\ref{fig:sketch_summand}).
We consider for definiteness three simple protocols for the pulsed magnetic gradient.
They have in common a zero net velocity shift (in order to close the loop
for the center-of-mass DoF) and their duration $2T_{1/2}$.
Moving for simplicity the moment of maximum splitting to $t = 0$, we take
for $-T_{1/2} \le t \le T_{1/2}$:
\begin{align}
a_0(t) & = \pm a_{\rm max} \qquad \text{(square profile, see inset Fig.\,\ref{fig:sketch_summand})}
\label{eq:simple-a0-model}
\\
a_1(t) & = a_{\rm max} [ - 1 + 6 (t/T_{1/2})^2 - 5(t/T_{1/2})^4 ]
\label{eq:simple-a1-model}
\\
a_2(t) & = - \frac{ a_{\rm max} }{ 2 }[ \cos( \pi t/T_{1/2} ) + \cos( 2 \pi t/T_{1/2} ) ]
\label{eq:simple-a2-model}
\end{align}
Their Fourier transforms have envelopes that scale with $1/\omega^{n+1}$ ($n = 0, 1, 2$):
\begin{align}
a_0(\omega, t) &=
\frac{ a_{\rm max} }{ \omega }
\left[ 2 \sin (\omega  T_{1/2}) - 4 \sin(\omega T_{1/2} / 2) \right]
\\
a_1(\omega, t) &=
- 16 a_{\rm max} T_{1/2} \left[
\frac{ \cos(\omega T_{1/2}) }{ (\omega T_{1/2})^2 }
\left( 1 - \frac{15}{(\omega T_{1/2})^2} \right)
\right.
\nonumber\\
& \qquad
\left. {}
-
\frac{ \sin(\omega T_{1/2}) }{ (\omega T_{1/2})^3 }
\left( 6 - \frac{15}{(\omega T_{1/2})^2} \right)
\right]
\\
a_2(\omega, t) &=
- 3 \pi^2 a_{\rm max} T_{1/2} \frac{\sin(\omega T_{1/2})}{(\omega T_{1/2})^3}
\nonumber\\
& \qquad
\times
\left( 1 - \frac{   \pi^2 }{ (\omega T_{1/2})^2 }\right)^{-1}
\left( 1 - \frac{ 4 \pi^2 }{ (\omega T_{1/2})^2 }\right)^{-1}
\label{eq:spectrum-acceleration}
\end{align}
%
For small \NOs, one expects that high frequency modes
should follow adiabatically the slowly varying acceleration and
return to their initial state. The values of the Fourier
transform $a_i(\omega_q, t)$ then estimate
how large is the \emph{deviation} from adiabaticity.
The protocol~2 shows the smallest results with
$a_2(\omega_q, t) \sim 1/\omega_q^{3}$ because
its acceleration has the smoothest (most adiabatic) onset.
The difference between the temporal profiles can be traced back to the experience
of a traveller, as either an old-fashioned or a modern train takes off.
As the particle increases in size, the phonon modes shift to lower frequencies and adiabatic following becomes less easier. By this mechanism, the interference contrast gets worse.

\begin{figure}[htbp]
\centerline{%
\includegraphics[width=0.47\textwidth]{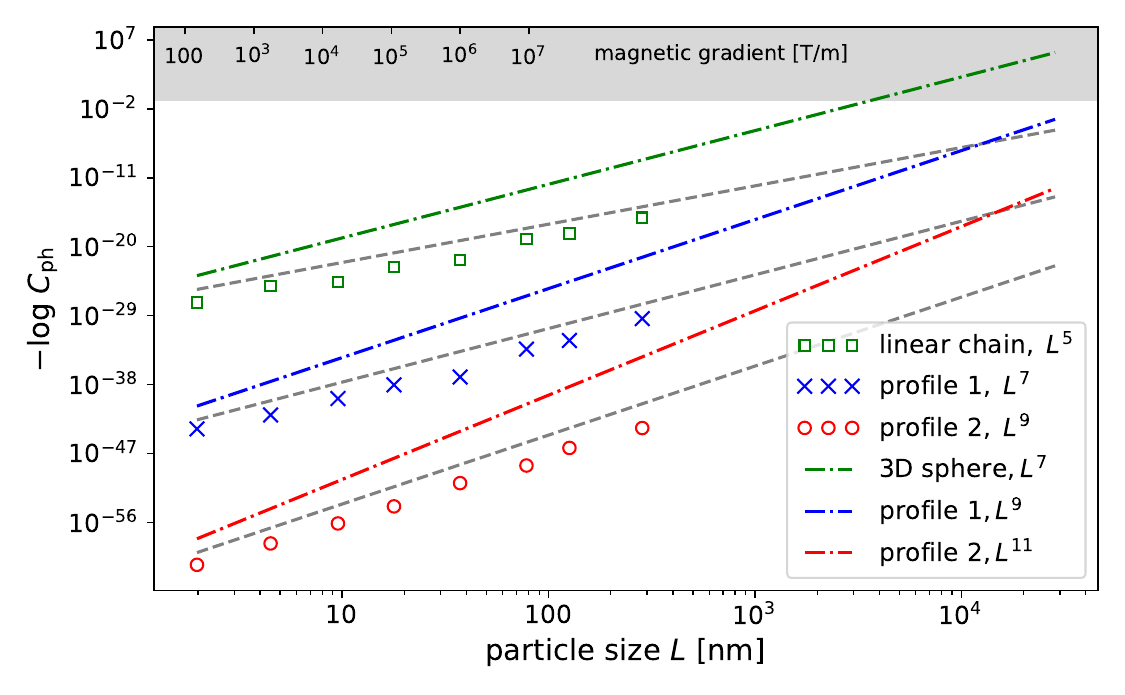}
}
\vspace*{-3ex}
\caption[]{%
Exponent in interference contrast $C_{\rm ph}$
for a one-dimensional chain (symbols and
gray dashed) and a three-dimensional, round object (dash-dotted).
$\log$ denotes the natural logarithm;
in the gray-shaded area, the contrast drops below $10\%$.
Lattice parameters for diamond at room temperature,
maximum acceleration kept fixed at $a_{\rm max} = 100\,{\rm m/s}^2$,
duration $2T_{1/2} = 60\,\mu{\rm s}$, spin position near the center,
(internal) temperature $T_{\rm ph} = 293\,{\rm K}$.
The data sets marked `linear chain' and `3D sphere' correspond to
the closed-loop acceleration protocol $a_0(t)$
[Eq.\,(\ref{eq:simple-a0-model})],
while `profile 1 and 2' correspond
to Eqs.\,(\ref{eq:simple-a1-model}--\ref{eq:simple-a2-model}), respectively.
The top scale gives the magnetic gradient required to achieve the
acceleration $a_{\rm max}$ for a 3D object, it scales with its volume $L^3$.
}
\label{fig:contrast-1D}
\end{figure}

The overall contrast reduction is very tiny, however, as can be
seen in Fig.\,\ref{fig:contrast-1D}. The symbols give
the sum in
Eq.\,(\ref{eq:result-phonon-contrast})
for one-dimensional \NOs of increasing length.
The overall scaling can be understood by focusing on the fundamental phonon mode,
i.e., the orthogonality for $\omega_1 = \pi c / L$.
We write $-\log C_{\rm ph} = f S$ with the prefactor
\begin{equation}
f = \frac{ (a_{\rm max} T_{1/2})^2 k_{\rm B} T_{\rm ph} M }{ \hbar^2 \omega_1^2 }
= \left( \frac{ a_{\rm max} T_{1/2} }{ \omega_1 \lambda_{\rm ph} }
\right)^2
\propto L^3 \qquad \text{(1D chain)}
\label{eq:scale-factor-f}
\end{equation}
where the phonon coherence `length' $\lambda_{\rm ph}$ (measuring actually a displacement) turns out to be equal to the thermal de Broglie wavelength for the chain's total mass $M \sim L$, but taken at the internal temperature $T_{\rm ph}$.
Adopting the approximation that the fundamental tone is in the
classical regime, $\hbar \omega_1 \ll k_B T_{\rm ph}$
(at room temperature, frequencies $< 6\,{\rm THz}$ or
size $L > 3\,{\rm nm}$),
the factor $f \sim T_{\rm ph}$ makes the scaling with the
phonon temperature explicit.
The other factor $S$ is a dimensionless sum
over the phonon spectrum and takes the form
\begin{align}
S &= \sum_q
\frac{ \beta \omega_1^2 }{ 2 \omega_q }
\coth\frac{ \beta \omega_q }{ 2 }
\cos^2[(s+\tfrac12)qa]
\left| \frac{ a(\omega_q, t) }{ a_{\rm max} T_{1/2} } \right|^2
\nonumber\\
& \simeq
\frac{ A_n }{ (\omega_1 T_{1/2} )^{2n+2} }
\sum_q
(\omega_1 / \omega_q)^{2n+4}
\propto L^{2n+2}
\label{eq:classical-estimate-C}
\end{align}
with $n = 0, 1, 2$.
In the second line, we took
$\cos^2[(s+\tfrac12)\pi a/L] \le 1$ as upper limit and applied an upper bound
for the Fourier spectra~(\ref{eq:spectrum-acceleration}),
leading to the numbers $A_0 = 36$, $A_1 = (16/\pi)^2$, $A_2 = 9$.
For an equidistant phonon spectrum,
the sum in the second line evaluates to
$\zeta(2n+4) \approx 1.08232, 1.01734, 1.00408$,
being dominated by its first term.

The resulting lower limit for $C_{\rm ph} = \exp( - f S )$
is shown in Fig.\,\ref{fig:contrast-1D}
as dashed gray lines, while
the symbols give the full sum computed numerically, showing good agreement.
By placing the impurity spin near the center of the \NO, the excitation
of the fundamental phonon is reduced, improving the contrast compared to Eq.\,(\ref{eq:classical-estimate-C}).
For the nano-diamond considered here, the temperature is such that the
upper part of the phonon spectrum is in the quantum regime,
$\hbar \omega_q > k_B T$, where $\coth(\beta\omega_q/2) \to 1$ stays
above its classical limit.
Due to the steep power laws, these modes make
a negligible contribution, however, and do not exceed the
estimate\,(\ref{eq:classical-estimate-C}).

The main message of this plot is that the actual values for the contrast
$C_{\rm ph}$ are extremely close to unity:
the interference contrast is \emph{not reduced at all} by the excitation of
phonon modes in a linear chain. In practice, the limits will be rather set
by the magnetic gradient (values given at the top of Fig.\,\ref{fig:contrast-1D})
and the spin coherence time (typically $\sim 200\,\mu{\rm s}$).\cite{Trusheim_2013} The case of a fixed force (rather than
acceleration) is considered in Fig.\,\ref{fig:contrast-3D-fixed-F} below.

We have checked that a mismatch of the final velocity does not qualitatively
change these results. For that, we considered models with adjusted coefficients
in such a way that the initial acceleration shows the same behaviour, but its
integral $\Delta v$ is nonzero.
It turns out that for any small particle and reasonably slow gradients
(in the sense given above, $c / 2L > 1/T_{1/2}$),
the high-frequency asymptote of the Fourier spectrum $a_1(\omega, 2T_{1/2})$
shows the same scaling, only its amplitude
is changed. For our purposes, the previously discussed closed-loop case $\Delta v = 0$
is thus sufficient.

\subsection{Three-dimensional particle}

The exact calculation of phonon modes for a three-dimensional assembly
of a few thousand atoms or more
becomes challenging. We sketch here the modifications that are needed with
respect to the linear chain. The wave vector $q$ becomes three-dimensional,
and a mode function like $u_q \cos[(n+\frac12) qa]$ becomes a vector-valued function
$u_{\bf q} {\bf f}_{\bf q}( {\bf r}_n )$ with mode amplitude $u_{\bf q}$. In the projection
formula Eq.\,(\ref{eq:mode-uq}), $N$ is now the total number of atoms in the \NO,
and in Eq.\,(\ref{eq:phonon-mode-q}), $M$ becomes the total mass. The Stern-Gerlach
force on the impurity spin (at ${\bf r}_s$)
defines a direction that selects an angular pattern
of emitted phonons via a scalar product with the local ``polarisation vector''
${\bf f}_{\bf q}( {\bf r}_s )$
of the phonon mode.
The shape of the \NO and its boundary conditions determine the allowed values
of $q$ within the Brillouin zone of the crystal structure. While an exact calculation
is possible (numerically) for small clusters, we focus in the following on the
region near the $\Gamma$ point with an approximately linear dispersion. The value
$q = 0$ corresponds again to the center-of-mass mode, and we are interested in the
smallest non-zero sound frequency $\omega_1$ (think of the fundamental pitch of
a musical instrument). One has
to allow for different values of the speed of sound $c$ (longitudinal and transverse).
In a \NO with cubic shape, the modes are separable, and
we have $\omega_1 = \pi c / L$ with the linear
size $L$. In a spherical object of diameter $D$, adopting a continuum model
and solving the Helmholtz equation
with a Neumann boundary condition, we find a dipole mode proportional
to the gradient of $\cos(\theta) \left[\sin(q r)/r^2 - q \cos(q r)/r \right]$
(a spherical Bessel function) at $\omega_1 = c q$
with $q \approx 4.1632/D$.
We note that experiments on resonances observed
with icosahedral clusters in quasi-crystalline materials
show good agreement with such a continuum analysis, even though the
cluster diameters are as small as $1\,{\rm nm}$.\cite{Duval_2005}
{A more complete analysis of the acoustic modes of a nano-sphere can be found in Ref.\,\onlinecite{GonzalezBallestero_2020b}.}

For a small particle (recall the typical limit $L < 30\,{\rm cm}$),
all phonon frequencies are way beyond the cutoff frequency
$1/T_{1/2}$ of the acceleration spectrum $a(\omega_{\bf q}, 2T_{1/2})$,
and the lowest phonon mode gives the dominant contribution.
Its second harmonic already contributes only a few percent,
depending on the protocol.
This being said, we get the following rough estimate for the interference
contrast for a small three-dimensional \NO:
\begin{align}
- \log C_{\rm ph} & \simeq
\frac{ M }{ 2 \hbar \omega_1 }
\coth\frac{ \beta \omega_1 }{ 2 }
|{\bf e}_z \cdot {\bf f}_1( {\bf r}_s )|^2
|a(\omega_1, t)|^2
\nonumber\\
& =
(\cdots)
\frac{ M k_{\rm B} T a_{\rm max}^2 }{ \hbar^2 \omega_1^6 T_{1/2}^2 }
=
(\cdots)
\frac{ (\Delta Z_{\rm max})^2 }{ \lambda_{\rm ph}^2 }
\left( \frac{ L }{ c T_{1/2} } \right)^6
\label{eq:result-phonon-contrast-0}
\end{align}
where ${\bf e}_z$ gives the direction of the Stern-Gerlach force, and
${\bf f}_1( {\bf r}_s )$ is the fundamental phonon mode (normalized
to unit maximum amplitude), evaluated at the position of the spin ${\bf r}_s$.
In the second line,
$(\cdots)$ is a numerical factor we expect to be of order
unity, we adopted protocol $a_1(t)$ and assumed $\hbar\omega_1 \ll k_{\rm B} T$.
The scaling with the linear dimension $L$ of the particle now gives the
exponent $L^{9}$ ($L^{7}$ for $a_0(t)$ and $L^{11}$ for $a_2(t)$),
see dash-dotted lines in Fig.\,\ref{fig:contrast-1D}.

The second form of Eq.\,(\ref{eq:result-phonon-contrast-0}) makes contact
to the seminal estimation of decoherence due to Zurek \cite{Zurek_2003b}:
the maximum spatial splitting (at half-loop)
$\Delta Z_{\rm max} \simeq a_{\rm max} T_{1/2}^2$
of the \NO wavepacket
is compared to the phonon coherence length $\lambda_{\rm ph}$.
This huge ratio would preclude any realistic contrast,
were it not compensated by the high power of the small ratio $L / c T_{1/2}$,
as long as the object size $L$ is smaller than the travelling distance of sound
during the closed loop.

\begin{figure}[htbp]
\centerline{%
\includegraphics[width=0.47\textwidth]{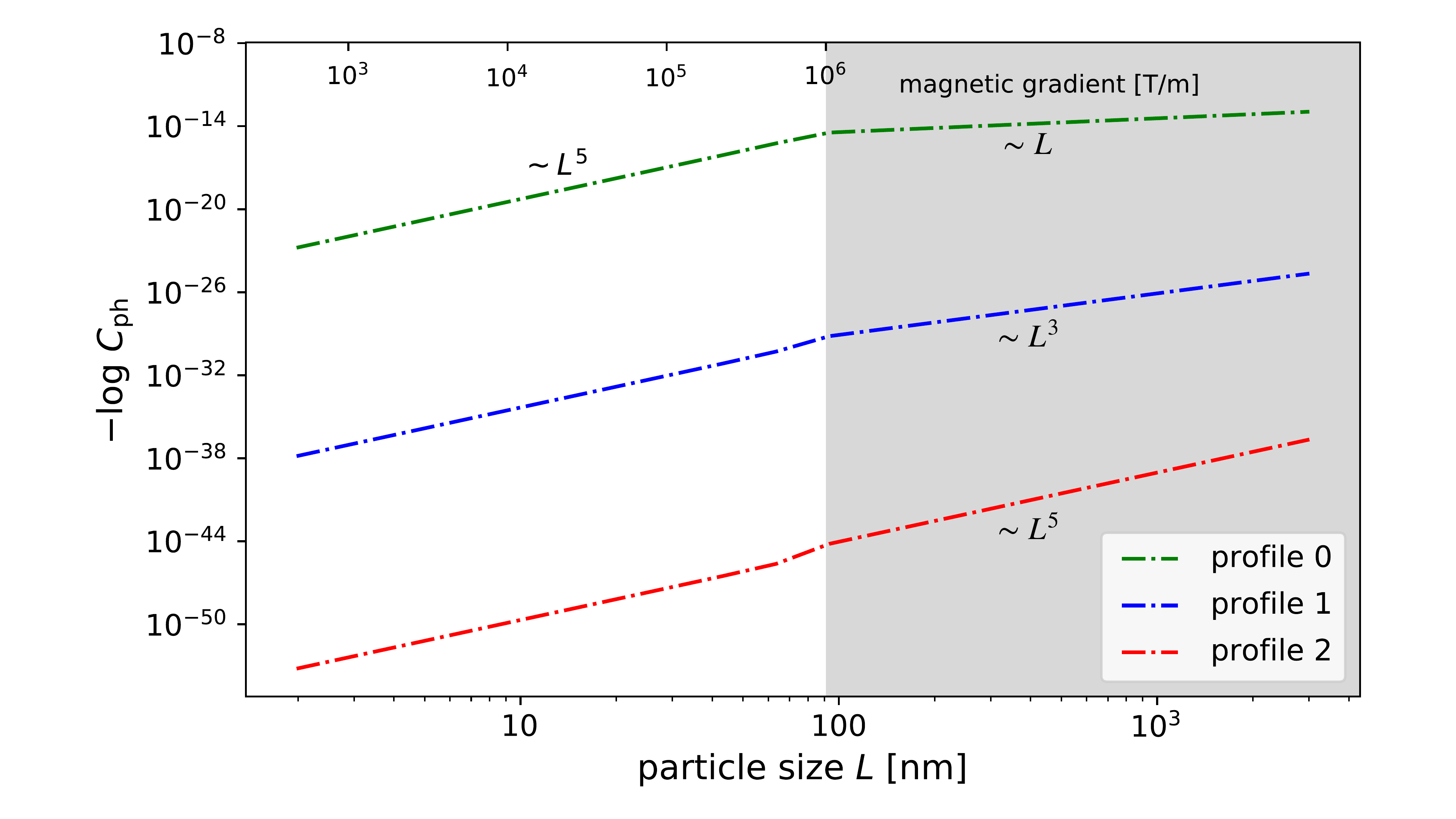}
}
\vspace*{-3ex}
\caption[]{%
Interference contrast for a three-dimensional particle.
Magnetic gradient (upper scale) and duration of the loop
are adjusted such that the maximum spatial splitting is a fixed fraction
($10\%$) of the \NO size, and the maximum velocity splitting is
$1\,{\rm mm/s}$. In the gray shaded area, the required magnetic gradient
and half-loop time are set to realistic maximum values ($10^{6}\,{\rm T/m}$
and $100\,\mu{\rm s}$ \cite{Trusheim_2013}) so that the targeted splitting
noted above is not achieved.
Other parameters as in Fig.\,\ref{fig:contrast-1D}.
The three data sets correspond to the
closed-loop protocols of
Eqs.\,(\ref{eq:simple-a0-model}--\ref{eq:simple-a2-model}), respectively,
\revised{they show the same scaling $\sim L^5$ with the particle diameter $L$ in the non-shaded area. The scaling laws differ in the gray area, as indicated.}
}
\label{fig:contrast-3D-fixed-F}
\end{figure}

In Fig.\,\ref{fig:contrast-3D-fixed-F}, the contrast is shown for a different
setting of parameters:
here, the magnetic gradient is such that the wavepacket splits to a
fixed fraction ($\sim 10\%$) of the \NO diameter. This would be a typical
requirement for a Stern-Gerlach interferometer with two particles that
interact via their mutual gravitational attraction, as suggested for
probing quantum gravity.\cite{Bose_2017, Marletto_2017}
We adjusted the protocol time
$T_{1/2}$ to get a fixed velocity splitting $\Delta v_{\rm max}$.
In the non-shaded area of the plot, the required values are below the limits
$10^6\,{\rm T/m}$ and $100\,\mu{\rm s}$ set by the current experimental
device in the Ben-Gurion University of the Negev group. (In the shaded area, the achieved splitting is
less than the targeted value.)
The contrast is again excellent, and comparing to Fig.\,\ref{fig:contrast-1D},
these settings permit to split somewhat larger particles. The
protocols $a_0$, $a_1$, and $a_2$ now differ by factors independent of the
particle size and scale all with the power $L^5$.
In fact, in this setting, the ratio $L/(c T_{1/2})$ turns
out to be constant ($\sim \Delta v_{\rm max}$ divided by the speed of sound),
as long as one stays below the upper limit to $T_{1/2}$.

\subsection{Further insight: macroscopic particle}

We finally consider the limit that the acoustic modes become dense on the
scale $2\pi/T_{1/2}$ of the acceleration spectrum. Although experiments will
be extremely challenging (size $L > 30\,{\rm cm}$), we include this case as a reference.

The sum over phonon wave vectors may be replaced by an integral (recall
the spacing $\Delta q = \pi /L$). Going directly to the three-dimensional
case, the exponent in the
contrast Eq.\,(\ref{eq:result-phonon-contrast}) becomes (average $|{\bf e}_z
\cdot {\bf f}_{\bf q}( {\bf r}_s )|^2 \approx \frac12$, $3$ acoustic branches)
\begin{align}
- \log C_{\rm ph} &\simeq
\frac{ 3 M L^3 }{ 2 \hbar }
\int\!\frac{ {\rm d}^3q }{ (2 \pi)^3 \omega_{\bf q} }
\coth\frac{ \beta \omega_{\bf q} }{ 2 }
|a(\omega_{\bf q}, t)|^2
\nonumber\\
& \simeq
\frac{ 3 M L^3 k_{\rm B} T }{ \pi \hbar^2 c^3  }
\int\!\frac{ {\rm d}\omega }{ 2 \pi }
|a(\omega, t)|^2
\label{eq:}
\end{align}
In the second line, we focused on the acoustic part of the dispersion relation
and took the high-temperature limit. This is a good approximation, since
the integration range is effectively
limited by the bandwidth $2\pi/T_{1/2}$ of the acceleration spectrum.
The integral gives, according to the Parseval-Plancherel formula:
\begin{equation}
\int\!\frac{ {\rm d}\omega }{ 2 \pi }
|a(\omega, 2T_{1/2})|^2
=
\int\limits_{-T_{1/2}}^{T_{1/2}}\!{\rm d}t\, a(t)^2
= C_n a_{\rm max}^2 T_{1/2}
\label{eq:}
\end{equation}
with $C_0 = 1$ for $a_0(t)$ and $C_1 = 256/315$, $C_2 = 1/2$.
Note that in this regime, the three protocols are essentially equivalent.
In terms of the phonon coherence length
$\lambda_{\rm ph}$,
the contrast becomes
\begin{equation}
- \log C_{\rm ph} \simeq
(\cdots)
\left( \frac{ \Delta Z_{\rm max} }{ \lambda_{\rm ph} } \right)^2
\left( \frac{ L }{ c T_{1/2} } \right)^3
\label{eq:}
\end{equation}
where again $\Delta Z_{\rm max} \simeq a_{\rm max} T^2_{1/2}$ and
$(\cdots)$ is a numerical coefficient of order one.
Since now the last factor is larger than unity (large particle), coherent splitting can only occur over distances smaller than $\lambda_{\rm ph} \ll 10^{-15}\,{\rm m}$. For the mass region stated in the beginning of this section, non-negligible splitting will thus be impossible.

It is interesting to note that when the acceleration is expressed by the
maximum force, $F_{\rm max} = M a_{\rm max}$,
the object mass $M$ drops out of this estimate. In terms of magnetic
gradient and protocol time, the requirement $C_{\rm ph} \ge 10\%$ gives the
inequality
\begin{equation}
\frac{ T_{1/2} }{ \mu{\rm s} }
\left( \frac{ b_{\rm max} }{ {\rm T/m} } \right)^2
\le (\cdots) 10^{15}
\frac{ [\varrho / ({\rm g/cm}^3)] [c/(10^3\,{\rm m/s})]^3
}{
(\mu / \mu_B)^2 (T_{\rm ph}/300\,{\rm K}) }
\label{eq:}
\end{equation}
This upper limit is larger than the parameters available in current experiments,
but only by a few orders of magnitude. The main challenge seems to be that
the corresponding forces on large objects are too small to split a wavepacket by a significant fraction of the object size (see
Fig.\,\ref{fig:contrast-3D-fixed-F}).

\section{Discussion and Conclusion}

In this work we examined the coherent splitting of a nano-object with an embedded single spin, in a full-loop Stern-Gerlach interferometer.
We addressed the question of whether the internal degrees of freedom (phonons) pose a problem, as they are excited by kicks in opposite directions for the two wavepackets. This may cause distinguishability to the point of orthogonality.

We looked at 1D and 3D models, assuming phonon normal modes in the harmonic approximation, and examined the scaling of the contrast reduction with the particle size at fixed acceleration or fixed maximal separation.
We looked at the coherence drop solely due to phonons, assuming that the closing (overlap) of other degrees of freedom (position, momentum, rotation) is perfect.
We took care to make use of realistic experimental numbers for the magnetic gradients and spin coherence time inside a nano-object.

As expected, we find that achieving complete overlap of the phonon state is impossible. However, the suppression of coherence for practical numbers is found to be  minimal. We examined several temporal profiles of the magnetic field and found that the smoother the profile (namely, slower onset of the magnetic field), the higher the eventual coherence. This is caused by an increasingly adiabatic behavior of the phonon modes. We also find that a non-zero temperature does not significantly alter the contrast. However, we find that once we go for higher masses, phonons indeed suppress the possibility of coherent interferometry, 
and this may indeed turn out to be a fundamental limit for creating massive superpositions in the future.

The final conclusion is therefore that phonon dynamics are not an inhibiting factor for a large range of parameters in nano-object Stern-Gerlach interferometry, making this method highly relevant for the fundamental experiments noted in the Introduction.

\smallskip

\paragraph*{Data Availability.}
The data that supports the findings of this study are available within the article.

\paragraph*{Conflict of Interest.}
The authors have no conflicts to disclose.

\revised{%
\paragraph*{Acknowledgments.}
This work was funded in part by the Israel Science Foundation (grant nos. 856/18, 1314/19, 3151/20, and 3470/21), by the Israel Innovation Authority within the QuantERA network (project LEMAQUME, grant no. 76974), 
and by the Deutsche Forschungsgemeinschaft through the DIP program (grant no. Fo 703/2-1).
}


\providecommand{\noopsort}[1]{}\providecommand{\singleletter}[1]{#1}%

\end{document}